\documentclass[prd,twocolumn,showpacs,preprintnumbers,amsmath,amssymb,superscriptaddress,floatfix,nofootinbib]{revtex4}
\usepackage{graphicx}
\usepackage{amsmath}
\usepackage{amsfonts}
\usepackage{amssymb}%
\usepackage{euscript}
\usepackage[dvips]{color}

\begin{document}
\renewcommand{\theenumi}{(\alph{enumi})}

\title{$\bar{B}^0$ and $\bar{B}^0_s$ decays into $J/\psi$ and $f_0(1370),~f_0(1710),~f_2(1270),~f'_2(1525),~K^*_2(1430)$}
\date{\today}
\author{Ju-Jun Xie}
\affiliation{Institute of Modern Physics, Chinese Academy of
Sciences, Lanzhou 730000, China} \affiliation{State Key Laboratory
of Theoretical Physics, Institute of Theoretical Physics, Chinese
Academy of Sciences, Beijing 100190, China}

\author{E.~Oset}
\affiliation{Institute of Modern Physics, Chinese Academy of
Sciences, Lanzhou 730000, China} \affiliation{Departamento de
F\'{\i}sica Te\'orica and IFIC, Centro Mixto Universidad de
Valencia-CSIC Institutos de Investigaci\'on de Paterna, Aptdo.
22085, 46071 Valencia, Spain}

\begin{abstract}

We make predictions for the ratios of branching fractions of ${\bar
B}^0$ and ${\bar B}^0_s$ decays into $J/\psi$ and the scalar mesons
$f_0(1370),~f_0(1710)$, or tensor mesons
$f_2(1270),~f'_2(1525),~K^*_2(1430)$. The theoretical approach is
based on results of chiral unitary theory where these resonances are
shown to be generated from the vector meson-vector meson
interaction. Eight independent ratios can be predicted and
comparison is made with the recent data on $\bar{B}^0_s$ decay into
$J/\psi f'_2(1525)$ versus the $\bar{B}^0_s$ decay into $J/\psi
f_2(1270)$.

\end{abstract}


\maketitle

\section{Introduction}

While there is a growing support for the low lying scalar mesons
$f_0(500)$, $f_0(980)$, $a_0(980)$, $\kappa(800)$ to be generated
dynamically from the interaction of pseudoscalar mesons, forming
some kind of composite meson meson
states~\cite{npa,ramonet,kaiser,markushin,juanito,rios}, the case of
the next set of scalar resonances at higher energies,
$f_0(1370),~f_0(1710)$, $K^*_0(1430)$ is more a question of debate.
So is the case of the tensor resonances
$f_2(1270),~f'_2(1525),~K^*_2(1430)$. Concerning the latter ones
there is some support for these resonances to be plain $q \bar q$
states belonging to a nonet of
$SU(3)$~\cite{Klempt:2007cp,Crede:2008vw}. The case of the scalar
resonances is more varied-the $f_0(1370)$ is also sometimes assumed
to be a $q \bar q$ state, although in
Refs.~\cite{Klempt:2007cp,Crede:2008vw} it is also suggested that it
could correspond to a $\rho \rho$ molecule based on phenomenological
properties (the decay widths into $\rho \rho$ and $\eta \eta$).
However, in Ref.~\cite{giacosa}, based on a study of decay
properties within the chiral linear sigma model, the $f_0(1370)$ is
suggested to be a $q \bar q$ state while the $f_0(1500)$ would
correspond to the glueball. In Ref.~\cite{Xia:2010zze} a study is
conducted of effects in some decays widths of assuming quarkonium,
tetraquark and gluonium components in the context of a nonlinear
chiral Lagrangian for the $f_0(500)$, $f_0(1370)$, and $f_0(1500)$.
There are even doubts about the existence of the $f_0(1370)$, but a
strong case in favor is made in Ref.~\cite{Bugg:2007ja}. In
Ref.~\cite{Fariborz:2006xq} the $f_0(500)$, $f_0(980)$ and
$f_0(1370)$ are assumed to be admixtures of two- and four-quark
components, with the $f_0(500)$ being dominantly a nonstrange
four-quark state, and the $f_0(980)$ and $f_0(1370)$ having a
dominant two-quark component. Similarly, $f_0(1500)$ and $f_0(1710)$
have considerable two- and four-quark admixtures, but in addition
they have a large glueball component. In Ref.~\cite{giacosados}
solutions in which the $f_0(1710)$ would be a glueball, while the
$f_0(1370)$ and $f_0(1500)$ are predominantly $q \bar q$ states, are
found likely. On the other hand in Ref.~\cite{migueloller} the
$f_0(1710)$ is advocated as a glueball, while the $f_0(1500)$ is
also assumed to have a large glueball component, while the
$f_0(1370)$ would correspond to a simple $q \bar q$ state. This is
just a sample or recent discussions on these issues; further
information and discussions can be found in the
reports~\cite{Klempt:2007cp,Crede:2008vw,Ochs:2013gi}.

On the other hand, a new perspective on these states has been
offered by the work of Ref.~\cite{raquel}, where the $f_0(1370)$ and
$f_2(1270)$ resonances were shown to be generated from the $\rho
\rho$ interaction provided by the local hidden gauge
Lagrangians~\cite{hidden1,hidden2,hidden4} implementing
unitarization. The work was extended to $SU(3)$ in
Ref.~\cite{gengvec} and 11 resonances were dynamically generated,
some of which were identified with the $f_0(1370)$, $f_0(1710)$,
$f_2(1270)$, $f'_2(1525)$ and $K^*_2(1430)$. The idea has been
tested successfully in a large number of reactions. In
Ref.~\cite{Nagahiro:2008um} the two-photon decay of the $f_0(1370)$
and $f_2(1270)$ were studied and good rates were obtained compared
with experiment. This latter work was extended in
Ref.~\cite{Branz:2009cv} to the study of the two photons and one
photon-one vector decays of the $f_0(1370)$, $f_2(1270)$,
$f_0(1710)$, $f'_2(1525)$, and $K^*_2(1430)$. In
Ref.~\cite{alberzou} a study of the $J/\psi \to \phi(\omega)
f_2(1270), ~f'_2(1525)$, and $J/\psi \to K^{*0}(892)
\bar{K}^{*0}_2(1430)$ decays was also carried out from that
perspective and good results were obtained. The radiative decay of
$J/\psi$ into $f_2(1270)$, $f'_2(1525)$, $f_0(1370)$ and $f_0(1710)$
was also studied in Ref.~\cite{Geng:2009iw} and good results were
obtained for the available experimental information. One also very
interesting repercussion of this perspective was the
reinterpretation in Ref.~\cite{MartinezTorres:2012du} of the peak
seen in the $\omega \phi$ distribution close to
threshold~\cite{Ablikim:2006dw} as a manifestation of the
$f_0(1710)$ resonance below the $\omega \phi$ threshold rather than
a signal for a new resonance. It is clear that the idea of the
nature of these states as vector meson-vector meson composite states
has undergone a scrutiny that no other model has undergone. Yet, the
permanent discussion of the issue demands that extra checks are done
with the different models for other observables, and in this sense
the weak decays that we exploit here bring a new source of valuable
information that should serve to test the different models. This is
the purpose of the present work.

In this work we present results for the weak decay of $\bar{B}^0$
and $\bar{B}^0_s$ decays into $J/\psi$ and $f_0(1370)$, $f_0(1710)$,
$f_2(1270)$, $f'_2(1525)$, $K^*_2(1430)$. The experimental results
show that the $\bar{B}^0_s$ has a pronounced peak for the decay into
$J/\psi f_0(980)$~\cite{Aaij:2011fx}, while no appreciable signal is
seen for the $f_0(500)$. These results have been also supported by
Belle~\cite{Li:2011pg}, CDF~\cite{Aaltonen:2011nk},
D0~\cite{Abazov:2011hv}, and again
LHCb~\cite{LHCb:2012ae,Aaij:2014emv} Collaborations. Conversely, in
Ref.~\cite{Aaij:2013zpt} the $\bar{B}^0$ into $J/\psi$ and $\pi^+
\pi^-$ is investigated and a clear signal is seen for the $f_0(500)$
production, while only a very small contribution from the $f_0(980)$
production is observed. These reactions have motivated theoretical
work, estimating rates of
production~\cite{Colangelo:2010bg,Wang:2009rc,Lu:2013jj} or trying
to extract the amount of $q \bar q$ or tetraquarks in the scalar
mesons~\cite{Stone:2013eaa}. Related theoretical work is also done
in Refs.~\cite{Wang:2009azc,robert,bruno,cheng,bruno2,lucio}. From
the perspective of the scalar mesons as being dynamically generated
from the meson-meson interaction, work was recently completed in
Ref.~\cite{weihong}, where the elementary mechanism is $J/\psi$
formation together with a $q \bar q$, which is hadronized to convert
it into a meson-meson pair. The resulting meson-meson pairs are
allowed to interact, using for this purpose the chiral unitary
approach~\cite{review}, and the desired final state is selected.
This interaction is known to generate dynamically the low lying
scalar mesons~\cite{npa,ramonet,kaiser,markushin,juanito,rios} and
then one has a mechanism to produce all these resonances in these
decays up to a global normalization constant. The agreement found
with experiment for the different decays modes in
Ref.~\cite{weihong} is remarkable. We shall follow a similar path
here, but taking into account the fact that the scalar and tensor
resonances discussed above, $f_0(1370)$, $f_0(1710)$, $f_2(1270)$,
$f'_2(1525)$, $K^*_2(1430)$, are now generated from the vector
meson-vector meson interaction.

\section{Formalism}

\begin{figure*}[htbp]
\begin{center}
\includegraphics[scale=0.8]{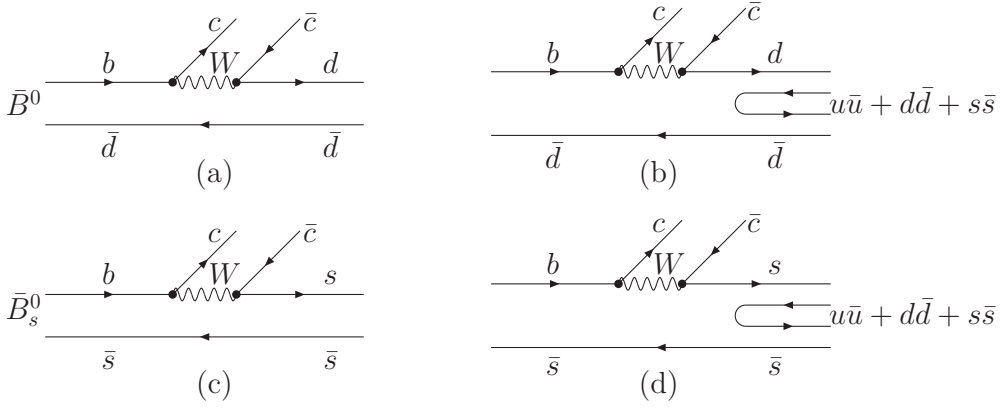}
\caption{ Basic diagrams for $\bar{B}^0$ and $\bar{B}^0_s$ decay
into $J/\psi$ and a $q\bar{q}$ pair [(a) and (c)], and hadronization
of the $q \bar{q}$ components [(b) and (d)].} \label{Fig:feyn}
\end{center}
\end{figure*}

Following the Ref.~\cite{Stone:2013eaa} we take the dominant
mechanism for the decay of $\bar{B}_0$ and $\bar{B}^0_s$ into a
$J/\psi$ and a $q\bar{q}$ pair. Posteriorly, this $q\bar{q}$ pair is
hadronized into vector meson-vector meson components, as depicted in
Fig.~\ref{Fig:feyn}.

The hadronization of the $q\bar{q}$ pair is done following the idea
of Ref.~\cite{alberzou}. The idea is depicted in Fig. 1. In the
$\bar B^0$ ($\bar B^0_s$) decays, a $c \bar c$ state is created
producing the $J/\psi$, together with a light $q \bar q$ pair, $d
\bar d$ for $\bar B^0$ and $s \bar s$ for $\bar B^0_s$ decays.  In
order to produce a meson which comes from vector-vector interaction,
this $q \bar q$ pair has to hadronize. We need four quarks to build
this new structure and this is accomplished by creating an extra
$\bar q  q$ combination with the quantum numbers of the vacuum,
which corresponds to a flavor structure $\bar u u+ \bar d d+ \bar s
s$. In order to formulate the correspondence between the hadronized
$q \bar q ( \bar u u+ \bar d d+ \bar s s)$ structure and that of the
vector-vector components one proceeds as follows. We introduce the
$\bar{q} q$ matrix $M$
\begin{equation}\label{eq:qqbarmatrix}
M=\left(
           \begin{array}{ccc}
             u\bar u & u \bar d & u\bar s \\
             d\bar u & d\bar d & d\bar s \\
             s\bar u & s\bar d & s\bar s \\
           \end{array}
         \right)
\end{equation}
which has the property
\begin{equation} \label{eq:MM}
M\cdot M=M \times (\bar u u + \bar d d + \bar s s).
\end{equation}

In this sense the hadronized $d\bar{d}$ and $s\bar{s}$ states in
Fig.~\ref{Fig:feyn} can be written as
\begin{eqnarray}
d\bar{d}(\bar u u + \bar d d + \bar s s) &=& (M\cdot M)_{22}, \\
s\bar{s}(\bar u u + \bar d d + \bar s s) &=& (M\cdot M)_{33} .
\end{eqnarray}

But now it is convenient to establish the relationship of these
hadronized components with the vector meson-vector meson components
associated to them. For this purpose we write the matrix $M$ of
Eq.~(\ref{eq:qqbarmatrix}) in terms of the nonet of vector mesons
\begin{equation}\label{eq:vectormatrix}
V = \!\! \left( \!\!
           \begin{array}{ccc}
             \frac{\sqrt{2}}{2}\rho^0 +  \frac{\sqrt{2}}{2} \omega  & \rho^+ & K^{*+} \\
             \rho^- & - \frac{\sqrt{2}}{2} \rho^0 +  \frac{\sqrt{2}}{2} \omega  & K^{*0} \\
            K^{*-} & \bar{K}^{*0} & \phi \\
           \end{array}
       \!\!  \right),
\end{equation}
and then we associate
\begin{eqnarray}
&& d\bar{d}(\bar u u + \bar d d + \bar s s) \equiv (V\cdot V)_{22} \nonumber \\
&& = \rho^-\rho^+ + \frac{1}{2}\rho^0\rho^0 + \frac{1}{2} \omega \omega -\rho^0 \omega + K^{*0}\bar{K}^{*0},  \label{eq:ddbarhadronization} \\
&& s\bar{s}(\bar u u + \bar d d + \bar s s) \equiv (V\cdot V)_{33}
\nonumber \\
&& = K^{*-}K^{*+} + K^{*0}\bar{K}^{*0} + \phi \phi .
\label{eq:ssbarhadronization}
\end{eqnarray}

\begin{figure*}[htbp]
\begin{center}
\includegraphics[scale=0.7]{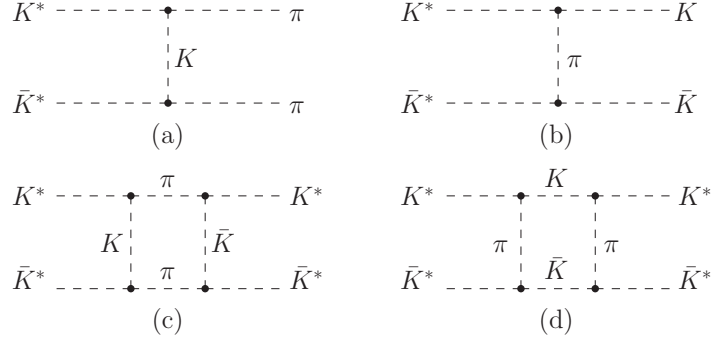}
\caption{ Decay mechanisms of $K^*\bar{K}^* + \pi\pi$, $K\bar{K}$
[(a) and (b)] and box diagrams considered in Ref.~\cite{gengvec} to
account for these decays [(c) and (d)].} \label{Fig:box}
\end{center}
\end{figure*}

In the study of Ref.~\cite{gengvec} a coupled channels unitary
approach was followed with the vector meson-vector meson states as
channels. However, the approach went further since, following the
dynamics of the local hidden gauge Lagrangians, a vector
meson-vector meson state can decay into two pseduoscalars, $PP$.
This is depicted in Figs.~\ref{Fig:box} (a) and 2(b). In
Ref.~\cite{gengvec} these decay channels are taken into account by
evaluating the box diagrams depicted in Figs.~\ref{Fig:box} (c) and
2(d), which are assimilated as a part, $\delta \tilde{V}$, of the
vector-vector interaction potential $\tilde{V}$. This guarantees
that the partial decay width into different channels could be taken
into account when determining masses and widths of the resonances
that the approach generates. However, the approach is not done
taking into account as coupled channels the $VV$ and $PP$
simultaneously. This means that one does not evaluate explicitly $VV
\to PP$ transition matrices. Although the partial decay widths into
$PP$ are well evaluated, the fact that one does not have the $VV \to
PP$ matrix elements forces us to take a path slightly different from
the one taken in Ref.~\cite{weihong} to deal with the low lying
scalar resonances, which are generated from the $PP$ interaction
solely. Hence, rather than evaluating amplitudes and mass
distributions for the pairs of pseudoscalars that are observed (the
resonances that we get are usually bound in the vector-vector
systems), we evaluate the amplitudes and rates for transition to the
resonance itself.

Since the information of the PDG~\cite{pdg} is usually given in
terms of rates for transition to specific resonances, the procedure
that we follow allows direct comparison with these experimental
magnitudes.

\begin{figure}[htbp]
\begin{center}
\includegraphics[scale=0.9]{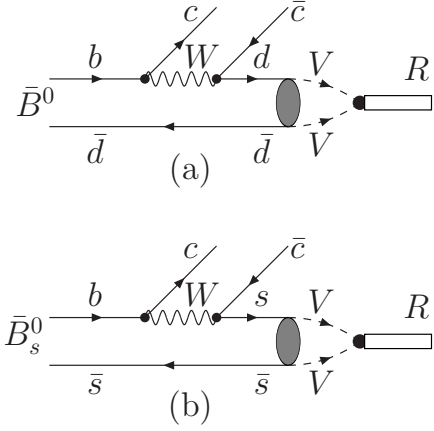}
\caption{ Mechanisms to generated the vector-vector resonances
through $VV$ rescattering. The dot of the vertex $RVV$ indicates the
coupling of the resonance to the different $VV$ components. (a) for
$\bar{B}^0$ decay, (b) for $\bar{B}^0_s$ decay.} \label{Fig:jpsiR}
\end{center}
\end{figure}

The vector-vector components of Eqs.~(\ref{eq:ddbarhadronization})
and (\ref{eq:ssbarhadronization}) are produced in a first step and
then they interact in coupled channels to produce finally the
desired resonance. This propagation is taken into account by means
of the two vector loop function $G_{VV}$, times the coupling of this
vector-vector component to the resonance. Since we wish to have the
resonance production and this is obtained through rescattering, the
mechanism for $J/\psi$ plus resonance production is depicted in
Fig.~\ref{Fig:jpsiR}.

The amplitudes for $J/\psi R$ production are then given by
\begin{eqnarray}
&& t(\bar{B}^0 \to J/\psi f_0) = \tilde{V}_P V_{cd} p_{J/\psi} {\rm
cos}\theta  (G_{\rho^- \rho^+} g_{\rho^-\rho^+, f_0} \nonumber \\
&& + \frac{1}{2}\frac{1}{2} G_{\rho^0\rho^0}g_{\rho^0\rho^0,f_0}
+ \frac{1}{2}\frac{1}{2}G_{\omega\omega}g_{\omega\omega,f_0} \nonumber \\
&& + G_{K^{*0}\bar{K}^{*0}}g_{K^{*0}\bar{K}^{*0},f_0}), \label{eq:bzerojpaifzero} \\
&& t(\bar{B}^0_s \to J/\psi f_0) = \tilde{V}_P V_{cs} p_{J/\psi}
{\rm cos}\theta (G_{K^{*0}\bar{K}^{*0}}g_{K^{*0}\bar{K}^{*0},f_0} \nonumber \\
&& + G_{K^{*-} K^{*+}} g_{K^{*-}K^{*+}, f_0}  +
\frac{1}{2}G_{\phi\phi}g_{\phi\phi,f_0} ) ,
\label{eq:bszerojpaifzero}
\end{eqnarray}
where $G_{VV}$ are the loop functions of two vector mesons that we
take from \cite{gengvec} and $g_{VV,f_0}$ the couplings of $f_0$ to
the pair of vectors $VV$, defined from the residues of the resonance
at the poles
\begin{eqnarray}
t_{ij} \simeq \frac{g_i g_j}{s-s_R},
\end{eqnarray}
with $t_{ij}$ the transition matrix from the channel $(VV)_i$ to
$(VV)_j$. These couplings are also tabulated in Ref.~\cite{gengvec}.
The formulas for the decay amplitudes to $J/\psi f_2$ are identical,
substituting $f_0$ by $f_2$ in the formulas and the factor
$\tilde{V}_P$ by a different one $\tilde{V}'_P$ suited for the
hadronization into a tensor. The magnitudes $\tilde{V}_P$ and
$\tilde{V}'_P$ represent the common factors to these different
amplitudes. In addition to the different weights of the several
vector-vector channels in Eqs.~(\ref{eq:bzerojpaifzero}) and
(\ref{eq:bszerojpaifzero}) for the $\bar B^0$ or $\bar B^0_s$ decays
into $J/\psi$ and the same resonance, one also has the weight of
different Cabibbo-Kobayashi-Maskawa (CKM) matrix element, $V_{cd}$
for $\bar B^0$ decay, and $V_{cs}$ for $\bar B^0_s$ decay. These
matrix elements are given by
\begin{eqnarray}
V_{cd} &=& - \rm{sin}\theta_c = -0.22534 , \\
V_{cs} &=& \rm{cos}\theta_c = 0.97427.
\end{eqnarray}

Note that in the formulas we include a factor $1/2$ in the $G$
functions for the $\rho^0 \rho^0$, $\omega \omega$, and $\phi \phi$
cases to account for the identity of the particles. The factor
$p_{J/\psi} {\rm cos}\theta$ is included there to account for a $p$
wave in the $J/\psi$ particle to match angular momentum in the $0^-
\to 1^-0^+$ transition. The $\rm{cos}\theta$ dependence is the
easiest one and we keep it, although it can be more complicated in
the presence of vector mesons, but this does not matter for the
ratios of rates. The factor $p_{J/\psi}$ can however play some role
due to the difference of mass between the different resonances.

\begin{figure}[htbp]
\begin{center}
\includegraphics[scale=0.8]{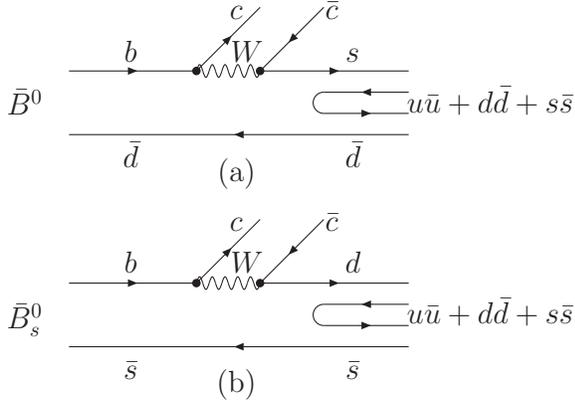}
\caption{ Mechanisms for $\bar{B}^0 \to J/\psi \bar{K}^{*}_2(1430)$
(a) and $\bar{B}^0_s \to J/\psi K^{*}_2(1430)$ (b).}
\label{Fig:Kstar}
\end{center}
\end{figure}

The case for $\bar{B}^0 \to J/\psi \bar{K}^{*}_2(1430)$ decay is
similar. The diagrams corresponding to Figs.~\ref{Fig:feyn} (b), (d)
are now written in Fig.~\ref{Fig:Kstar}.

In analogy to Eqs. (\ref{eq:ddbarhadronization}),
(\ref{eq:ssbarhadronization}) we now have
\begin{eqnarray} \label{eq:kstar}
&& s\bar{d}(u\bar{u} + d\bar{d} + s\bar{s}) \equiv (V\cdot V)_{32} \nonumber \\
&& = K^{*-} \rho^+ + \bar{K}^{*0} (-\frac{ \rho^0}{\sqrt{2}} + \frac{\omega}{\sqrt{2}} )  + \bar{K}^{*0} \phi , \\
&& d\bar{s}(u\bar{u} + d\bar{d} + s\bar{s}) \equiv (V\cdot V)_{23}
\nonumber \\
&& = \rho^- K^{*+} + (-\frac{ \rho^0}{\sqrt{2}} +
\frac{\omega}{\sqrt{2}} ) K^{*0} + K^{*0} \phi ,
\end{eqnarray}
and the amplitudes for production of $J/\psi \bar{K}^*_2(1430)$ will
be given by
\begin{eqnarray}
&& t(\bar{B}^0 \to J/\psi \bar{K}^*_2) = \tilde{V}'_P p_{J/\psi}
{\rm cos}\theta V_{cs} \nonumber \\
&& ( G_{K^{*-} \rho^+} g_{K^{*-}\rho^+, \bar{K}^*_2}
 - \frac{1}{\sqrt{2}} G_{\bar{K}^{*0} \rho^0} g_{\bar{K}^{*0}
\rho^0, \bar{K}^*_2} \nonumber \\
&& + \frac{1}{\sqrt{2}} G_{\bar{K}^{*0} \omega}g_{\bar{K}^{*0}
\omega, \bar{K}^*_2} + G_{\bar{K}^{*0} \phi}g_{\bar{K}^{*0} \phi,
\bar{K}^*_2} ), \label{eq:bzerojpsikstar} \\
&& t(\bar{B}^0_s \to J/\psi K^*_2) = \tilde{V}'_P p_{J/\psi}
{\rm cos}\theta V_{cd} \nonumber \\
&& ( G_{K^{*+} \rho^-}g_{K^{*+}\rho^-, K^*_2}
 - \frac{1}{\sqrt{2}} G_{\bar{K}^{*0} \rho^0} g_{\bar{K}^{*0}
\rho^0, K^*_2} \nonumber \\
&& + \frac{1}{\sqrt{2}} G_{\bar{K}^{*0} \omega}g_{\bar{K}^{*0}
\omega, K^*_2} \!\! + \! G_{\bar{K}^{*0} \phi}g_{\bar{K}^{*0} \phi,
K^*_2} ) . \label{eq:bszerojpsikstar}
\end{eqnarray}

One more step is needed since the couplings in
Eqs.~(\ref{eq:bzerojpaifzero}), (\ref{eq:bszerojpaifzero}),
(\ref{eq:bzerojpsikstar}), (\ref{eq:bszerojpsikstar}) are given in
charge basis while in the work of Ref.~\cite{gengvec} they are given
in isospin basis. For this we recall that in the unitary
normalization used in Ref.~\cite{gengvec} for convenience to deal
with identical particles one has
\begin{eqnarray}
&& |\rho\rho, I=0> = - \frac{1}{\sqrt{6}} (\rho^-\rho^+ +
\rho^0\rho^0  + \rho^+ \rho^-), \\
&& |K^*\bar{K}^*, I=0> = - \frac{1}{\sqrt{8}} ( K^{*-}K^{*+} +
K^{*0}
\bar{K}^{*0} \nonumber \\
&& ~~~~~~~~~~~~~~~~~~~~~~~~ + K^{*+} K^{*-} + \bar{K}^{*0}K^{*0} ) , \\
&& |\omega \omega, I=0> = \frac{1}{\sqrt{2}} \omega \omega , \\
&& |\phi \phi, I=0> = \frac{1}{\sqrt{2}} \phi \phi .
\end{eqnarray}

Then we find
\begin{eqnarray}
&& t(\bar{B}^0 \to J/\psi f_0) = \tilde{V}_P p_{J/\psi} {\rm
cos}\theta V_{cd}  ( -\frac{5\sqrt{6}}{12} G_{\rho \rho} g_{\rho
\rho, f_0} \nonumber \\
&& + \frac{\sqrt{2}}{4} G_{\omega \omega}g_{\omega \omega , f_0} -
\frac{\sqrt{2}}{2} G_{K^* \bar{K}^*} g_{K^* \bar{K}^*, f_0} ),
\\
&& t(\bar{B}^0_s \to J/\psi f_0) = \tilde{V}_P p_{J/\psi} {\rm
cos}\theta V_{cs} \nonumber \\
&& ( -\sqrt{2} G_{K^* \bar{K}^*} g_{K^* \bar{K}^*, f_0} +
\frac{\sqrt{2}}{2} G_{\phi \phi} g_{\phi \phi , f_0} ), \\
&& t(\bar{B}^0 \to J/\psi \bar{K}^{*}_2) = \tilde{V}'_P p_{J/\psi}
{\rm cos}\theta V_{cs}  ( -\frac{\sqrt{6}}{2} G_{\rho \bar{K}^*}
g_{\rho \bar{K}^*, \bar{K}^*_2} \nonumber \\
&& + \frac{\sqrt{2}}{2} G_{\omega \bar{K}^*} g_{\omega \bar{K}^* ,
\bar{K}^{*}_2} + G_{\phi \bar{K}^*} g_{\phi \bar{K}^*,
\bar{K}^{*}_2} ),
\\
&& t(\bar{B}^0_s \to J/\psi K^*_2) = \tilde{V}'_P p_{J/\psi} {\rm
cos}\theta V_{cd}  ( -\frac{\sqrt{6}}{2} G_{\rho K^*} g_{\rho K^*,
K^*_2} \nonumber \\
&& + \frac{\sqrt{2}}{2} G_{\omega K^*} g_{\omega K^* , K^*_2} +
G_{\phi K^*} g_{\phi K^*, K^*_2} ),
\end{eqnarray}
and for $\bar{B}^0 (\bar{B}^0_s) \to J/\psi f_2$ the same as for
$f_0$, but changing $\tilde{V}_P$ by $\tilde{V}'_P$. Note that
$\tilde{V}'_P$ is then common to the decays into $f_2$ and
$\bar{K}^*_2$.

The width for these decays will be given by
\begin{eqnarray}
\Gamma = \frac{1}{8\pi M^2_{\bar{B}}} |t|^2 p_{J/\psi} ,
\end{eqnarray}
with
\begin{eqnarray}
p_{J/\psi} = \frac{\lambda(M^2_{\bar{B}}, M^2_{J/\psi}, M^2_R)}{2
M_{\bar{B}}},
\end{eqnarray}
with $M_R$ the resonance mass, and in $|t|^2$ we include the factor
$1/3$ for the integral of ${\rm cos}\theta$, which cancels in all
ratios that we will study.

The information on couplings and values of the $G$ functions,
together with uncertainties, is given in Table V of
Ref.~\cite{alberzou} and Table I of Ref.~\cite{Geng:2009iw}. The
errors for the scalar mesons production are taken from
Ref.~\cite{Geng:2009iw}.

\section{Results}

In the PDG we find branching fractions for $\bar{B}^0_s \to J/\psi
f_0(1370)$~\cite{LHCb:2012ae}, $\bar{B}^0_s \to J/\psi
f_2(1270)$~\cite{LHCb:2012ae}, and $\bar{B}^0_s \to J/\psi
f'_2(1525)$~\cite{Aaij:2011ac}. We can calculate ten independent
rates and we have two unknown normalization constants $\tilde{V}_P$
and $\tilde{V}'_P$. As a consequence we can provide eight
independent ratios parameter free. From the present experimental
branching fractions we can only get one ratio for the $\bar{B}^0_s
\to J/\psi f_2(1270) [f'_2(1525)]$. There is only one piece of data
for the scalars, but we should also note that the data for
$\bar{B}^0_s \to J/\psi f_0(1370)$ in Ref.~\cite{LHCb:2012ae} and in
the PDG, in a more recent paper~\cite{Aaij:2014emv} is claimed to
correspond to the $f_0(1500)$ resonance. Similar ambiguities stem
from the analysis of Ref.~\cite{LHCb:2012ad}.

The branching fractions for $f_2(1270)$~\cite{LHCb:2012ae} and
$f'_2(1525)$~\cite{Aaij:2011ac} of the PDG are
\begin{eqnarray}
&& {\mathcal B}  (\bar{B}^0_s \to J/\psi f_2(1270)) = (10^{+5}_{-4})
\times \!\! 10^{-7} , \\
&& {\mathcal B}  (\bar{B}^0_s \to J/\psi f'_2(1525)) = (2.6
^{+0.9}_{-0.6})  \times \!\! 10^{-4}.
\end{eqnarray}

However, the datum for ${\mathcal B}(\bar{B}^0_s \to J/\psi
f_2(1270))$ of the PDG is based on the contribution of only one
helicity component $\lambda = 0$, while $\lambda = \pm 1$ contribute
in similar amounts.

This decay has been further reviewed in Ref.~\cite{Aaij:2014emv} and
taking into account the contribution of the different helicities a
new number is now provided,~\footnote{From discussions with S. Stone
and L. Zhang. This new number has been submitted to the PDG by the
authors of Ref.~\cite{Aaij:2014emv} and will appear in the next
update of the PDG.}
\begin{eqnarray}
{\mathcal B}(\bar{B}^0_s \to J/\psi f_2(1270)) = \!\! (3.0 ^{+1.2}
_{-1.0}) \! \times  \! \! 10^{-6}, \label{br1525}
\end{eqnarray}
which is about three times larger than the one already reported in
the PDG.

We present our results in Table~\ref{ratios} for the eight ratios
that we predict, defined as,
\begin{eqnarray}
R_1 &=& \frac{\Gamma [ \bar{B}^0 \to J/\psi
f_0(1370)]}{\Gamma [ \bar{B}^0 \to J/\psi f_0(1710)]} , \\
R_2 &=&  \frac{\Gamma [ \bar{B}^0 \to
J/\psi f_2(1270)]}{\Gamma [ \bar{B}^0 \to J/\psi f'_2(1525)]} , \\
R_3 &=&  \frac{\Gamma [ \bar{B}^0 \to
J/\psi f_2(1270)]}{\Gamma [ \bar{B}^0 \to J/\psi \bar{K}^*_2(1430)]} , \\
R_4 &=&  \frac{\Gamma [ \bar{B}^0 \to
J/\psi f_0(1710)]}{\Gamma [ \bar{B}^0_s \to J/\psi f_0(1710)]} , \\
R_5 &=&  \frac{\Gamma [ \bar{B}^0 \to J/\psi
f_2(1270)]}{\Gamma [ \bar{B}^0_s \to J/\psi f_2(1270)]} , \\
R_6 &=&  \frac{\Gamma [ \bar{B}^0_s \to
J/\psi f_0(1370)]}{\Gamma [ \bar{B}^0_s \to J/\psi f_0(1710)]}, \\
R_7 &=&  \frac{\Gamma [ \bar{B}^0_s \to
J/\psi f_2(1270)]}{\Gamma [ \bar{B}^0_s \to J/\psi f'_2(1525)]} , \\
R_8 &=&  \frac{\Gamma [ \bar{B}^0_s \to J/\psi f_2(1270)]}{\Gamma [
\bar{B}^0_s \to J/\psi K^*_2(1430)]} .
\end{eqnarray}

\begin{table}[htbp]
\begin{center}
\caption{Ratios of $\bar{B}^0$ and $\bar{B}^0_s$ decays.}
\begin{tabular}{|c|c|c|}
\hline \hline Ratios  & Theory & Experiment \\
\hline $R_1 $ & $6.2 \pm 1.6$ & $--$ \\ \hline $R_2 $  & $13.4 \pm
6.7$ &$--$
\\ \hline
$R_3 $  & $(3.0 \pm 1.5) \times 10^{-2}$ & $--$ \\
\hline
$R_4 $  & $(7.7 \pm 1.9) \times 10^{-3}$ & $--$ \\
\hline
$R_5$   & $(6.4 \pm 3.2) \times 10^{-1}$ & $--$ \\
\hline
$R_6 $  & $(1.1 \pm 0.3) \times 10^{-2}$ & $--$ \\
\hline
$R_7 $  & $(8.4 \pm 4.6) \times 10^{-2}$ & $(1.0 \sim 3.8) \times 10^{-2}$ \\
\hline
$R_8 $  & $(8.2 \pm 4.1) \times 10^{-1}$ & $--$ \\
\hline \hline
\end{tabular} \label{ratios}
\end{center}
\end{table}

Note that the different ratios predicted vary in a range of
$10^{-3}$, which means a big range, such that even a qualitative
level comparison with future experiments would be very valuable
concerning the nature of the states as vector vector molecules, on
which the numbers of the Tables are based.

The errors are evaluated in quadrature from the errors in
Refs.~\cite{alberzou,Geng:2009iw}. In the case of $R_7$, where we
can compare with the experiment, we put the band of experimental
values for the ratio to show that the theoretical results and the
experiment just overlap within errors.

From our perspective it is easy to understand the small ratio of
these decay rates. The $f_2(1270)$ in Ref.~\cite{gengvec} is
essentially a $\rho \rho$ molecule while the $f'_2(1525)$ couples
mostly to $K^* \bar{K}^*$. If one looks at
Eq.~(\ref{eq:bszerojpaifzero}) one can see that the $\bar{B}^0_s \to
J/\psi f_0(f_2)$ proceeds via the $K^* \bar{K}^*$ and $\phi \phi$
channels, hence, the $f_2(1270)$ with small couplings to
$K^*\bar{K^*}$ and $\phi \phi$ is largely suppressed, while the
$f'_2(1525)$ is largely favoured.

One should take into account that the rate of Eq.~(\ref{br1525}) is
one of the smallest rates reported in the PDG. These numbers come
from an elaborate partial wave analysis that, although rather stable
against different assumptions, is not free of ambiguities. In this
context, the agreement of theory with experiment in the only case
that we can compare is very encouraging and shows the potential that
the measurement of the other ratios has in learning about the nature
of the set of resonances on which we have reported. This discussion
should serve to encourage further experimental analysis in this
direction.

\section{Conclusions}

In this paper we have studied the decay of  $B^0$ and $B^0_s$  into
$J/\psi$ and one of the resonances $f_0(1370)$, $f_0(1710)$,
$f_2(1270)$, $f'_2(1525)$, $K^*_2(1430)$, which are generated
dynamically from the interaction of vector mesons. The approach
followed is rather simple and very predictive. We isolate the
dominant mechanisms for the elementary decay of the $B$ into the
$J/\psi$ and a $q \bar q$ component. This latter one is hadronized,
giving rise to two vector mesons which are allowed to interact in
coupled channels with a unitary approach, with the input obtained
from the local hidden gauge approach, which extends chiral symmetry
to the realm of the vectors. The approach allows us to get ten
independent decay rates and we have two unknown factors in the
theory. They are eliminated to give eight independent ratios of
rates, which appear parameter free in the theory. We could only
compare with one of the smallest ratios, the one between the $B^0_s$
into $J/\psi f_2(1270)$ and $B^0_s$ into $J/\psi f'_2(1525)$, and
the agreement was good within errors. This small ratio has reasons
purely dynamical in our theory. Indeed, we could see that the decay
selected a $s \bar s$ pair that upon hadronization gets converted
into $K^* \bar K^*$ and $\phi \phi$. On the other hand, in the
underlying theory of these states as vector-vector molecules, the
$f_2(1270)$ couples essentially to $\rho \rho$, while the
$f'_2(1525)$ couples basically to $K^* \bar K^*$. Then it comes
naturally that the $f_2(1270)$ is largely suppressed, while the
$f'_2(1525)$ is clearly favored.

The potential of the ratios predicted to tell us about the dynamics
of vector interaction and the nature of the resonances discussed
here is great.  This should serve to encourage further measurements
and analysis of data. At the same time, to advance on the issue of
the nature of resonances, it would be most advisable that other
groups, with other theories, also make predictions for these rates
that allow one to make comparisons and advance in our understanding
of the nature of hadronic resonances.

\section*{Acknowledgments}

One of us, E. O., wishes to acknowledge support from the Chinese
Academy of Science (CAS) in the Program of Visiting Professorship
for Senior International Scientists. We would like to thank Diego
Milanes for useful discussions and motivation to do this work, and
S. Stone and L. Zhang for enlightening discussions concerning the
experiments. This work is partly supported by the Spanish Ministerio
de Economia y Competitividad and European FEDER funds under the
contract number FIS2011-28853-C02-01 and FIS2011-28853-C02-02, and
the Generalitat Valenciana in the program Prometeo II-2014/068. We
acknowledge the support of the European Community-Research
Infrastructure Integrating Activity Study of Strongly Interacting
Matter (acronym HadronPhysics3, Grant Agreement n. 283286) under the
Seventh Framework Programme of EU. This work is also partly
supported by the National Natural Science Foundation of China under
Grant No. 11105126. The Project Sponsored by the Scientific Research
Foundation for the Returned Overseas Chinese Scholars, State
Education Ministry.

\end{document}